\documentclass[12pt,a4paper]{article}

\usepackage[english]{babel}
\usepackage{amsmath,amssymb,graphicx,psfrag}
\usepackage{subfigure}

\numberwithin{equation}{section}

\title{Domino tilings and the six-vertex\\ model at its free fermion point}
\author{Patrik L.\ Ferrari and Herbert Spohn \\[6pt]
{\normalsize Technische Universit\"at M\"unchen}\\
{\normalsize Zentrum Mathematik and Physik Department}\\
{\normalsize e-mails: ferrari@ma.tum.de, spohn@ma.tum.de}}
\date{16th May 2006}

\begin{document}
\sloppy
\maketitle

\begin{abstract}
At the free-fermion point, the six-vertex model with domain wall boundary conditions (DWBC) can be related to the Aztec diamond, a domino tiling problem. We study the mapping on the level of complete statistics for general domains and boundary conditions. This is obtained by associating to both models a set of non-intersecting lines in the Lindstr\"{o}m-Gessel-Viennot (LGV) scheme. One of the consequence for DWBC is that the boundaries of the ordered phases are described by the Airy process in the thermodynamic limit.
\end{abstract}

\section{Introduction}
Statistical mechanics models with short range interactions have a free energy per unit volume which is independent of boundary conditions. The six-vertex model seems to be an exception, at first glance. If one imposes domain wall boundary conditions (DWBC), the free energy differs from the one with periodic boundary conditions
even in the infinite volume limit \cite{KZJ00}.

\begin{figure}[t!]
\begin{center}
\psfrag{w1}[c]{$\omega_1$}
\psfrag{w2}[c]{$\omega_2$}
\psfrag{w3}[c]{$\omega_3$}
\psfrag{w4}[c]{$\omega_4$}
\psfrag{w5}[c]{$\omega_5$}
\psfrag{w6}[c]{$\omega_6$}
\includegraphics[height=4cm]{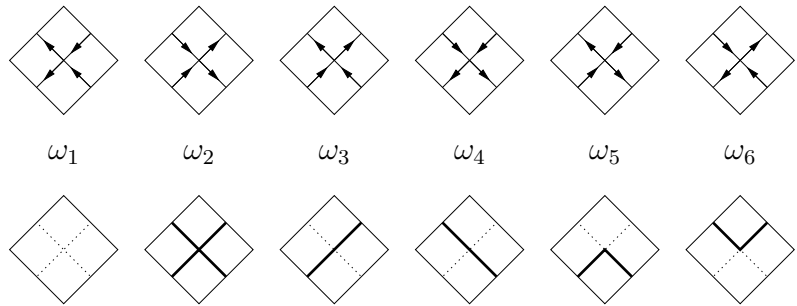}
\caption{The basic vertices of the six-vertex model (first row). By drawing only the NE and SE arrows one obtains the tiles in the second row.}\label{FigWeights}
\end{center}
\end{figure}

\begin{figure}[t!]
\begin{center}
\subfigure[\hspace{-0.8em}]{
\psfrag{S}[l][t]{$P_S$}
\includegraphics[height=5.8cm]{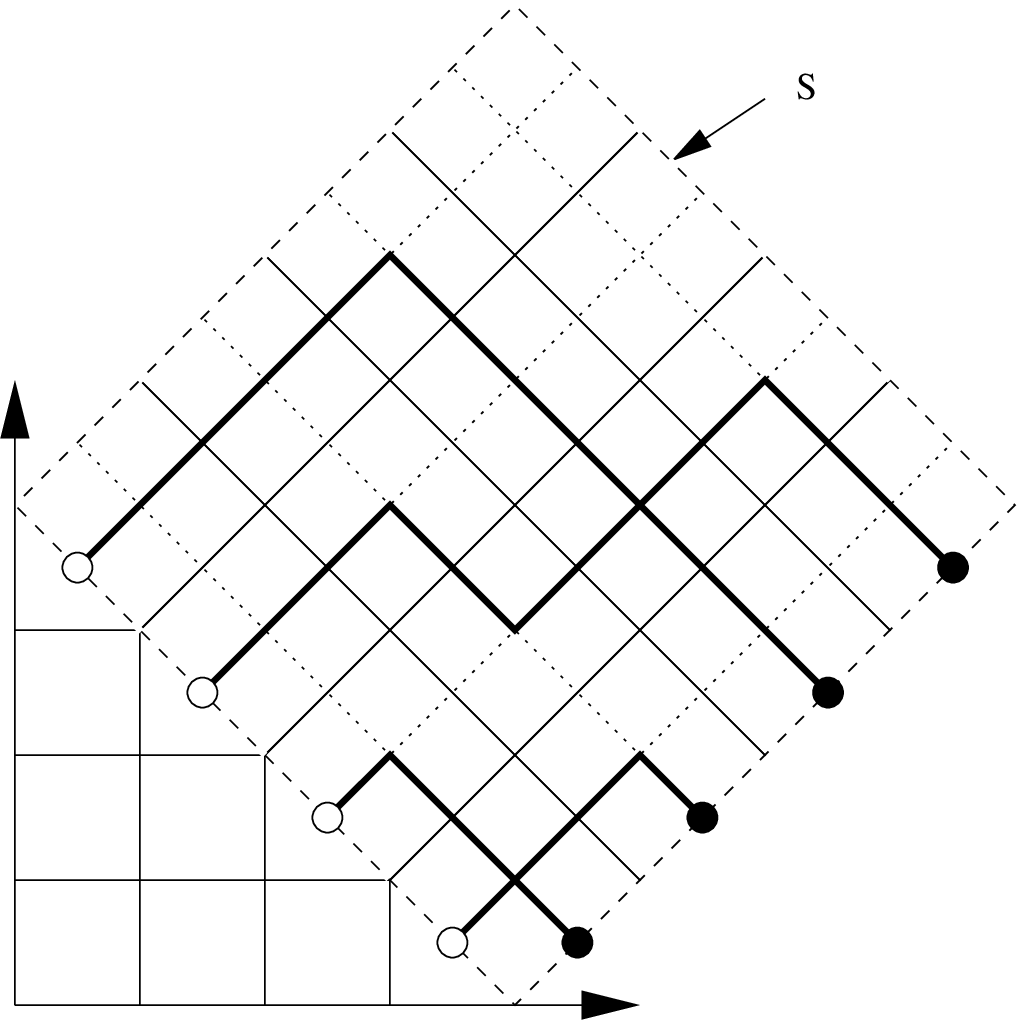}}
\hfill
\subfigure[\hspace{-0.8em}]{
\psfrag{PD}[l][t]{$P_D$}
\includegraphics[height=5.8cm]{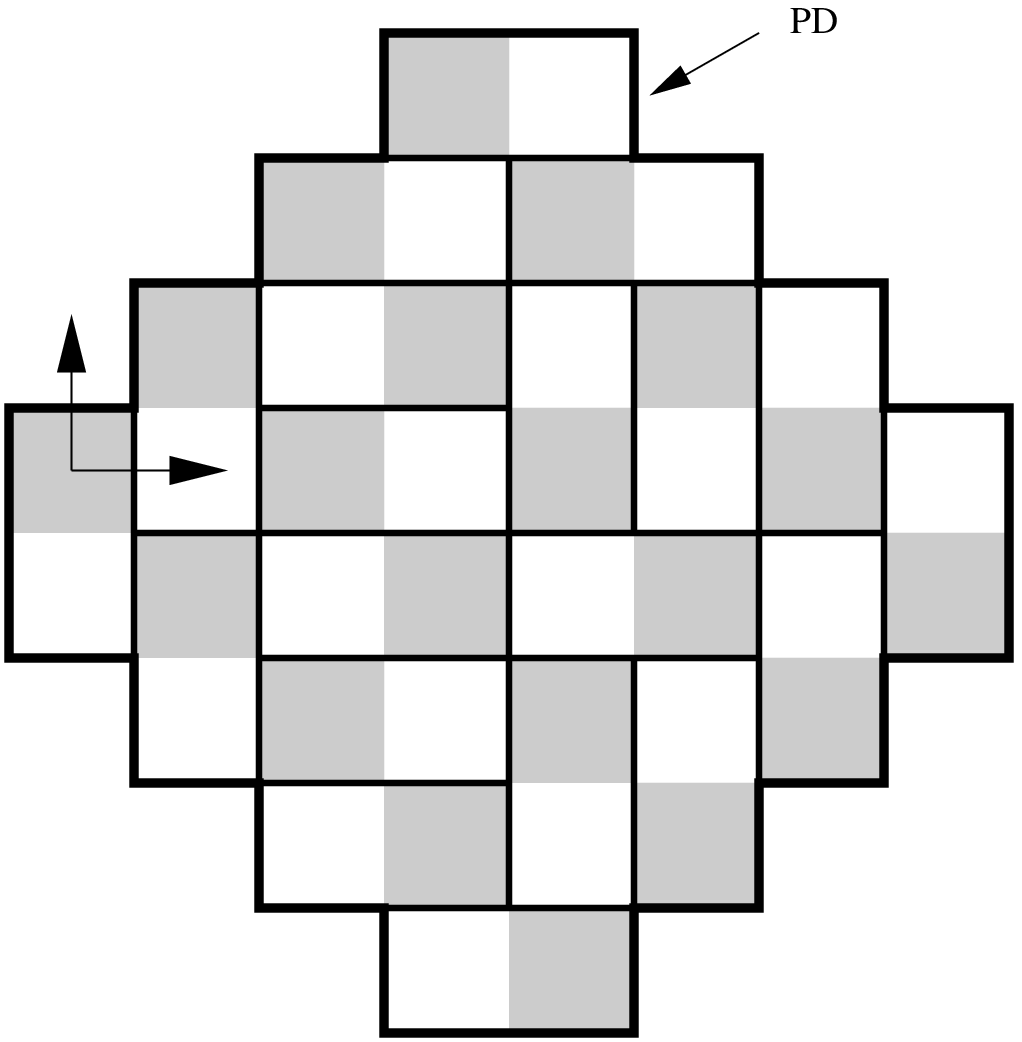}}
\caption{(a) A configuration of the 6-vertex model with DWBC. The lines start at the white dots and end at the black dots. (b) A configuration of the Aztec diamond.}\label{FigDWBCaztec}
\end{center}
\end{figure}

To explain more precisely we recall the six allowed vertices and their weight, see Figure~\ref{FigWeights}.
For better visualization only NE and SE oriented lines are drawn. The reason for the rotation by $\pi/4$, as compared to the usual arrows on $\mathbb{Z}^2$, will become apparent later. We want to think of the vertices in Figure~\ref{FigWeights} as tiles and the equilibrium statistical mechanics as a tiling problem. Let us then consider a domain $S$ and prescribe the tiles along its boundary $\partial S$, for an example see Figure~\ref{FigDWBCaztec}(a).
An admissible tiling is such that the lines are never broken and start, resp.~end, only at the boundary $\partial S$. In this language, DWBC correspond to a square-shaped domain such that the lines start regularly spaced at the SW edge and end regularly spaced at the SE edge, as in Figure~\ref{FigDWBCaztec}(a). The corresponding free energy is computed in \cite{KZJ00}, for finite size corrections see \cite{BF05,ZJ00}.

In the context of DWBC it has been noted that the six-vertex model maps to another, very widely studied tiling problem, namely the Aztec diamond, see Figure~\ref{FigDWBCaztec}(b). There one has given a diamond shaped domain in $\mathbb{Z}^2$ of size $2N\times 2N$, which has to be tiled with $2\times 1$, resp.~$1\times 2$, dominos. If every allowed tiling has equal weight, for large $N$ there is a central disordered zone of linear size $N$, which is bordered by four regular zones also of linear size $N$ and located at the four corners~\cite{CEP96}. Thus the limit state for $N\to \infty$ ``phase'' segregates. In contrast, for periodic boundary conditions the system is homogeneously disordered, which physically is the reason behind the boundary dependent free energies.

The Aztec diamond has been analyzed in great detail by means of determinantal processes. In particular one knows the precise statistics of the line dividing the ordered from the disordered zone \cite{Jo03}. The determinantal property suggests that equal weight in the domino tiling corresponds to the DWBC six-vertex model at its free fermion point. The aim of our contribution is to study the correspondence between domino tilings and the six-vertex model for general domains and boundary conditions, of course including DWBC, on the level of the complete statistics. With this information one readily translates the known statistical properties of the Aztec diamond to the DWBC six-vertex model at its free fermion point.

To give an outline: we introduce the domino tiling and the general six-vertex model for arbitrary domains and boundary conditions. We explain the Lindstr\"{o}m-Gessel-Viennot (LGV) scheme for nonintersecting paths on directed graphs and show in which sense the six-vertex model at the free fermion point is a special case of the general LGV set-up. Thereby we achieve directly the one-to-many mapping from six-vertex tilings to domino tilings.

\section{Six-vertex model with fixed boundary conditions}
In the usual set-up, at every vertex there are two incoming and two outgoing arrows (the so-called \emph{ice rule}). Thereby one obtains six distinct vertices with Boltzmann weights $\omega_1,\ldots,\omega_6$, see Figure~\ref{FigWeights}. Equivalently, one may think of a tiling problem with six distinct tiles by drawing only the North-East and the South-East arrows. A six-vertex configuration consists in a tiling such that the arrows of the vertices match amongst themselves and with the arrows prescribed at the boundary of the domain. In terms of the lines of Figure~\ref{FigWeights}, each configuration of the six-vertex model can be represented by a set of continuous lines which necessarily end at the boundary of the domain.

In this paper we mainly consider the case where the boundary is a rectangle. Later, we also discuss more general boundary conditions. So, let the border of the region we want to tile be a rectangle $P_S$ composed by segments $(\pm 1,\pm 1)$ between sites of $\mathbb{Z}^2$, see Figure~\ref{FigDWBCaztec}(a). The tiled region $S$ is the one enclosed by the rectangle $P_S$. Define the set $\partial S=P_S\cap (\mathbb{Z}+1/2)^2$, i.e., the mid-points of the basic segments forming the rectangle $P_S$. \emph{Fixed boundary conditions} means to define the points of $\partial S$ which are start-, resp.\ end-points of a line. In Figure~\ref{FigDWBCaztec}(a) the start points are the white dots, the end points the black ones. These points are on sides which can be oriented in four different directions as defined by the direction of the outer normal vector. The white dots lie on the SW or NW sides of $S$, the black ones on the SE or NE sides.

The six-vertex model with fixed boundary conditions has \emph{two free parameters}. Every elementary tile contains up to 2 line segments of length $1/\sqrt{2}$ in either the NE ($\diagup$) or SE ($\diagdown$) direction. Denote by $N_i$ the number of vertices of type $i$ (the ones with weight $\omega_i$) in a given configuration. There are four equations reducing the internal free parameters (the weights) from 6 to 2. Namely, there are some constants $C_1,C_2,C_3,C_4$ such that
\begin{enumerate}
\item[(1)] $N_1+N_2+N_3+N_4+N_5+N_6 = C_1$,
\item[(2)] $2N_2+N_5+N_6+2N_3= \textrm{number of }\diagup = C_2$,
\item[(3)] $2N_2+N_5+N_6+2N_4= \textrm{number of }\diagdown = C_3$,
\item[(4)] $N_5-N_6= \textrm{number of }\diagup\hspace{-1pt}\diagdown-\textrm{number of }\diagdown\hspace{-1pt}\diagup = C_4$.
\end{enumerate}
The first equation says that the total number of vertices is fixed by $S$. The second and third equation are a direct consequence of the fact that the start- and end-points are fixed. Finally, the fourth equation holds because given a possible set of lines, one can find any other set of lines by a subsequent shift of the corners $\diagup\hspace{-1pt}\diagdown$ and $\diagdown\hspace{-1pt}\diagup$. During this procedure it is not possible to increase (or decrease) the number of $\diagdown\hspace{-1pt}\diagup$ by one without forcing the same change to the number of $\diagup\hspace{-1pt}\diagdown$.

The weight of a configuration is
\begin{equation}
\prod_{i=1}^6 (\omega_i)^{N_i}
\end{equation}
and by using the above relations it can be rewritten as
\begin{equation}
C \tilde \omega_2^{N_2} \tilde \omega_5^{N_5}
\end{equation}
with $C=\omega_1^{(C_1-(C_3+C_4)/2)}\omega_3^{(C_2+C_4)/2}\omega_4^{(C_3+C_4)/2}\omega_6^{C_4}$, \mbox{$\tilde \omega_2=\omega_1\omega_2/\omega_3\omega_4$}, and \mbox{$\tilde \omega_5=\omega_5\omega_6/\omega_3\omega_4$}. Therefore, w.l.o.g., we can set
\begin{equation}
\omega_1=\omega_3=\omega_4=\omega_6=1
\end{equation}
and keep $\omega_2$, $\omega_5$ as \emph{free internal parameters}. This choice turns out to be convenient for the mapping.

To a tiling one can associate a height function on $S$ as follows. Fix a reference height equal to zero at some point on $P_S\setminus \partial S$. The height along the loop $P_S$ is increased (resp.\ decreased) by one unit when a white or black dot on $\partial S$ is crossed while going to NE or NW direction (resp.\ SE or SW). Inside $S$ the lines of the tiling are the boundaries of domains with the same height. In computing the surface free-energy, there are four free parameters since two are internal parameters and two correspond to chemical potentials determining the average slope, in accordance with our previous argument.

In the parameter space of the six-vertex model there is a codimension one manifold which is determined by the equation
\begin{equation}\label{eqFreeFermion}
\omega_1\omega_2+\omega_3\omega_4=\omega_5\omega_6
\end{equation}
and is called \emph{free-fermion point}.
Therefore, in case of fixed boundary conditions, there is only a single internal free parameter. Let us set
\begin{equation}
\omega_2=\alpha, \quad \omega_5=1+\alpha.
\end{equation}

The terminology ``free-fermion'' refers to a transfer matrix which is the exponential of a quadratic fermion hamiltonian. This roughly corresponds to line ensembles on a graph conditioned to be non-intersecting (no points in common). If we set $\alpha=0$, thus $\omega_2=0$, then the free-fermion condition (\ref{eqFreeFermion}) is automatically satisfied, and the corresponding line ensemble fits straightforwardly into the so-called Lindstr\"om-Gessel-Viennot (LGV) scheme to be explained below. For the case $\alpha>0$ this is no longer the case and Brak and Owczarek~\cite{BOw99} approach the problem by introducing ``osculating paths''. Actually, this is not needed and, as we will explain, the general free-fermion case can still be represented in a simple LGV scheme.

\section{Aztec diamond and domino tiling}
A graph closely related to the one presented here appears also in the domino tiling of a checkerboard region. A particular example is the Aztec diamond~\cite{Jo03}. The Aztec diamond consists in a diamond-shaped region $D$, see Figure~\ref{FigDWBCaztec}(b). The black sites are on even-even and odd-odd sites, the remaining sites are white. As for the six-vertex model, we can associate a simple polygonal loop $P_D$ enclosing all the $1\times 1$ squares centered on the elements of $D$. It is a loop which joins the sites of $(\mathbb{Z}+1/2)^2$ by segments of length $1$. We define the boundary points of $D$ by $\partial D=P_D\cap (\mathbb{Z}+1/2)^2$. For the Aztec diamond of parameter $N$, the set $D=D_N$ is given by
\begin{equation}
D_N=D_{N,w}\cup D_{N,b},
\end{equation}
where $D_{N,w}$ and $D_{N,b}$ are the white and black sites of the Aztec diamond,
\begin{eqnarray}
D_{N,b}&=&\{(x=m+n,y=n-m), m=0,\ldots,N,n=0,\ldots,N-1 \}, \\
D_{N,w}&=&\{(x=m+n,y=n-m-1),m=0,\ldots,N-1,n=0,\ldots,N\}.\nonumber
\end{eqnarray}

\begin{figure}[t!]
\begin{center}
\psfrag{N}[c]{$N$}
\psfrag{S}[c]{$S$}
\psfrag{E}[c]{$E$}
\psfrag{W}[c]{$W$}
\psfrag{wN}[c]{$w_N=1$}
\psfrag{wS}[c]{$w_S=\alpha$}
\psfrag{wE}[c]{$w_E=1$}
\psfrag{wW}[c]{$w_W=1$}
\includegraphics[height=4cm]{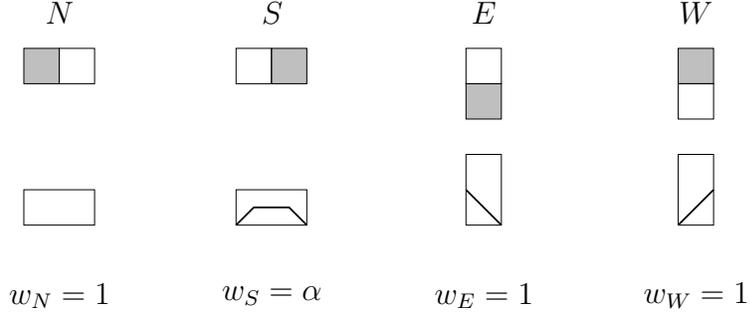}
\caption{Line configurations for the tiling (above) and the line configurations (below). The weight of $S$ is $\alpha$, the weights of $N$, $E$, $W$ are $1$.}\label{FigDominos}
\end{center}
\end{figure}

A tiling of the Aztec diamond (the region $D_N$) consists in taking $N(N+1)$ dominos ($2\times 1$ or $1\times 2$ tiles) to cover every site of $D_N$ (without overlapping). Depending on the direction of the dominos, vertical or horizonal, and their relative positions with respect to the black and white sites, one can distinguish 4 types of elementary tiles, called North, South, East, and West dominos, see Figure~\ref{FigDominos}.

To each of the 4 types of domino one assigns a Boltzmann weight. The Aztec diamond has \emph{one free parameter}. In fact, let $N_a$ be the number of dominos of type $a\in\{N,S,E,W\}$ in a configuration of the Aztec diamond. Then there are 3 equations reducing the free parameters from 4 to 1, namely
\begin{eqnarray}
N(N+1)&=&N_N+N_S+N_E+N_W,\nonumber \\
N_N&=&N_S,\\
N_E&=&N_W.\nonumber
\end{eqnarray}
Therefore the Aztec diamond has only one free-parameter. For easy comparison with the 6-vertex model, we set the weights of the dominos as follows
\begin{equation}
w_N=w_E=w_W=1,\quad w_S=\alpha.
\end{equation}

To each configuration of the Aztec diamond one can associate a non-intersecting set of lines. This is obtained by replacing the 4 types of dominos in the first row of Figure~\ref{FigDominos} with the lines as in the second row of the same illustration. As we will see in the next section, such line ensembles can also be associated to the 6-vertex model at its free-fermion point.

\section{Lindstr\"om-Gessel-Viennot scheme}
Let us first recall the Lindstr\"om-Gessel-Viennot scheme, see~\cite{Ste90} for details, and then apply it to the free-fermion six-vertex model.

Gessel and Viennot start with a graph $(V,E)$ consisting of vertices $V$ and directed edges $E$. The graph has \emph{no loops}. A path $P$ is a sequence of consecutive vertices joined by directed edges. $\mathcal{P}(u,v)$ is the set of all paths starting at $u\in V$ and ending at $v\in V$. The paths $P$ and $P'$ intersect, if they have a common vertex. Every edge carries a weight $w(e)$. The weight of a path $P$ is
\begin{equation}
w(P)=\prod_{e\in P\cap E}w(e)
\end{equation}
and we set
\begin{equation}
W(u,v)=\sum_{P\in \mathcal{P}(u,v)}w(P).
\end{equation}
We now consider an $r$-tuple $\vec{u}=\{u_1,\ldots,u_r\}$ of start-points and an $r$-tuple $\vec{v}=\{v_1,\ldots,v_r\}$ of end points. Let $\mathcal{P}_0(\vec{u},\vec{v})$ be the set of all
\emph{non-intersecting} $r$-tuple of paths from $\vec{u}$ to $\vec{v}$. $\vec{u}$ and $\vec{v}$ have to be \emph{compatible}, which means that any $r$-tuple of paths in $\mathcal{P}_0(\vec{u},\vec{v})$ necessarily connects $u_j$ to $v_j$ for $j=1,\ldots,r$. Then the weight of $\mathcal{P}_0(\vec{u},\vec{v})$ is given by
\begin{equation}
w(\mathcal{P}_0(\vec{u},\vec{v}))=\det \big(W(u_i,v_j)\big)_{1\leq i,j\leq r}.
\end{equation}
\begin{figure}[t!]
\begin{center}
\psfrag{a1}[c]{$1$}
\psfrag{a2}[c]{$1$}
\psfrag{a3}[c]{$1$}
\psfrag{a4}[c]{$1$}
\psfrag{a5}[c]{$\alpha$}
\includegraphics[height=3cm]{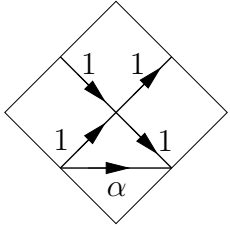}
\caption{The LGV elementary cell of the free-fermion six-vertex model.}\label{FigLGV}
\end{center}
\end{figure}

\begin{figure}[t!]
\begin{center}
\psfrag{w1}[c]{$\omega_1=1$}
\psfrag{w2}[c]{$\omega_2=\alpha$}
\psfrag{w3}[c]{$\omega_3=1$}
\psfrag{w4}[c]{$\omega_4=1$}
\psfrag{w5}[c]{$\omega_5=1+\alpha$}
\psfrag{w6}[c]{$\omega_6=1$}
\psfrag{1}[c]{$1$}
\psfrag{a}[c]{$\alpha$}
\includegraphics[height=5cm]{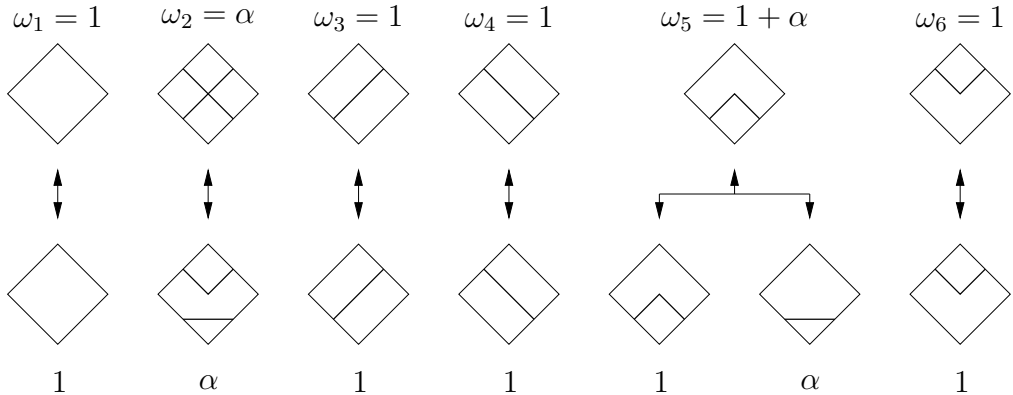}
\caption{Mapping from the 6-vertex configurations to LGV line configurations. The first row contains the 6-vertex elementary tiles, the second row the LGV line configurations. Notice that the map from the 6-vertex lines to the LGV lines is not one-to-one. We also give the weight of the respective elementary tiles.}\label{FigLGVmap}
\end{center}
\end{figure}

For the six-vertex model at its free fermion point, the LGV graph is build up from the elementary cells of Figure~\ref{FigLGV}. Thereby, the weights $\omega_i$'s are as required, see also Figure~\ref{FigLGVmap}. In fact, the only correspondence which is not one-to-one is for the weight $\omega_5=1+\alpha$ because there are two possible paths, one passing by the center of the cell of weight $1$ and the second following the edge with weight $\alpha$.
Therefore, in terms of non-intersecting line ensembles, we have a degeneracy. More precisely, to 6-vertex configurations with $N_5$ vertices of type 5, there are $2^{N_5}$ different LGV line configurations. However, these are only local modifications of each other and any observable not distinguishing the precise path inside an elementary cell remains unaffected.

\section{Connection between domino and six-vertex models}
\begin{figure}[t!]
\begin{center}
\subfigure[\hspace{-0.8em}]{
\psfrag{S}[l][t]{$P_S$}
\includegraphics[height=5.8cm]{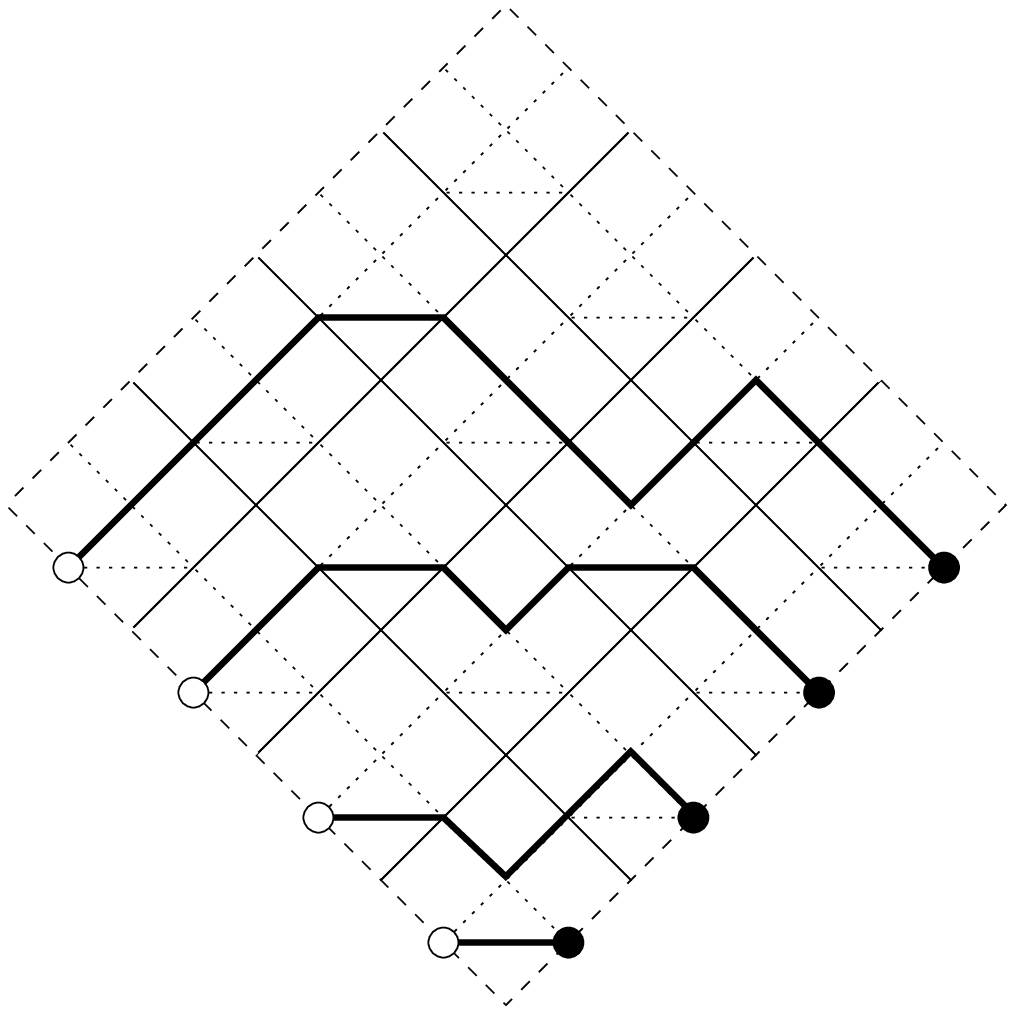}}
\hfill
\subfigure[\hspace{-0.8em}]{
\includegraphics[height=5.8cm]{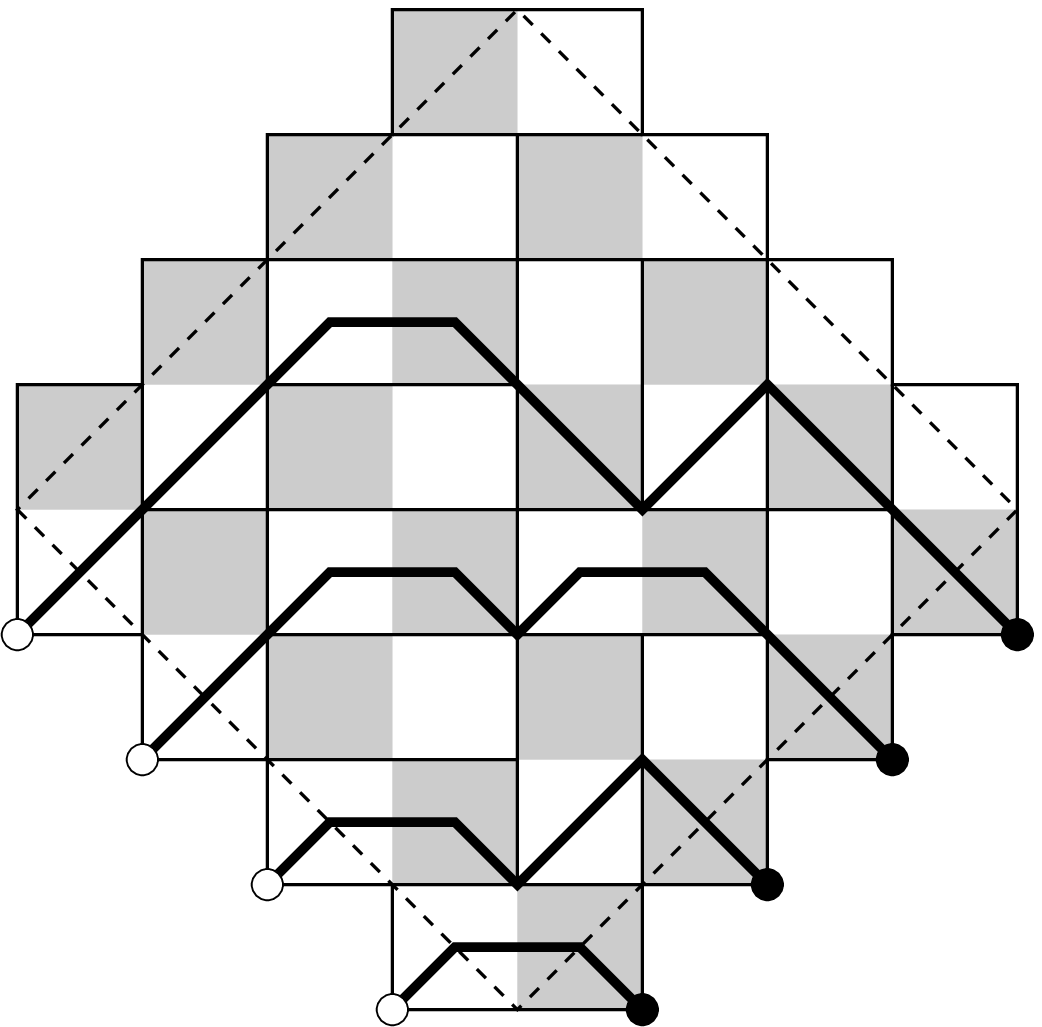}}
\caption{(a) One of the possible line-configuration of the 6-vertex model with DWBC of Figure~\ref{FigDWBCaztec}(a). The dotted lines are the edges of the subjacent directed graph of the LGV construction. (b) The corresponding Aztec diamond.}\label{FigDWBCaztecLines}
\end{center}
\end{figure}

The connection between the six-vertex model and domino tilings can be seen most directly in terms of line ensembles. Since the mapping from a six-vertex configuration to its associate set of lines is one-to-many, also the mapping to domino tilings will be one-to-many. In Figure~\ref{FigDWBCaztecLines}(a) there is one of the $2^5$ (since $N_5=5$) possible set of non-intersecting lines associated to the six-vertex configuration of Figure~\ref{FigDWBCaztec}(a). Figure~\ref{FigDWBCaztecLines}(b) is the corresponding configuration of the Aztec diamond with the same set of non-intersecting lines.

\subsubsection*{Six-vertex $\to$ domino tilings.}
Given a \emph{fixed boundary condition} for the six-vertex model on $\partial S$, we associate the fixed boundary conditions for domino tiling, i.e., we define the domain $D$ to be tiled with dominos and the height along the boundary $P_D$.

The idea is that the graph LGV supporting the non-intersecting lines remains, up to \emph{deterministic} local modifications, unchanged by the mapping to domino tilings. The LGV graph on the black and white sites of the domino tiling are
\begin{center}
\includegraphics[height=1cm]{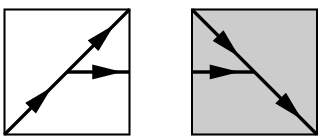}
\end{center}

\begin{figure}[t!]
\begin{center}
\psfrag{I1}[b][]{$I_1$}
\psfrag{I2}[b][]{$I_2$}
\psfrag{I3}[b][]{$I_3$}
\psfrag{I4}[b][]{$I_4$}
\psfrag{E1}[b][]{$E_1$}
\psfrag{E2}[b][]{$E_2$}
\psfrag{E3}[b][]{$E_3$}
\psfrag{E4}[b][]{$E_4$}
\includegraphics[width=\textwidth]{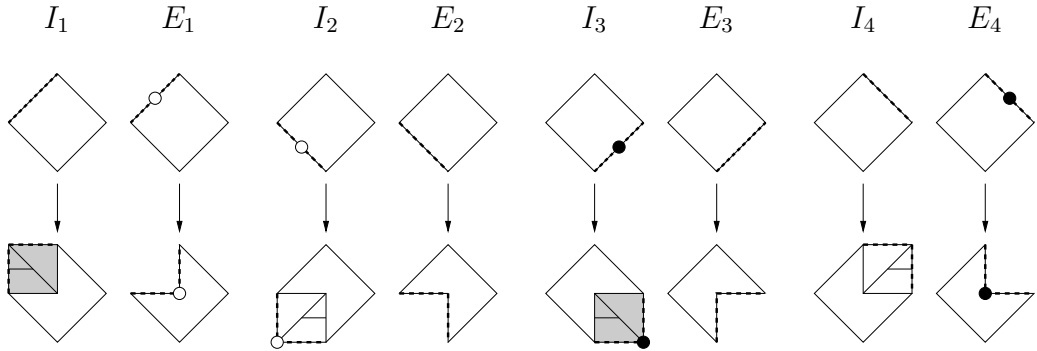}
\caption{The 8 possible boundary configurations of a cell of the six vertex model (first row) and its corresponding boundary configuration for domino tiling. The dashed lines represents the boundaries.}\label{FigBC}
\end{center}
\end{figure}

The region tiled by the six-vertex model and the one of the domino tiling cannot be identical. We have to decide whether the squares centered at the boundary $\partial S$ are to be tiled or not. This is achieved by the mapping represented in Figure~\ref{FigBC}.
The mapping for the cases $E_k$, $k=1,\ldots,4$, follows from the fact that, according to the lines for domino tiling introduced in Figure~\ref{FigDominos}, the deleted squares would be part of dominos with a second square outside the tiling region. On the other hand, the squares for the cases $I_k$, $k=1,\ldots,4$, are part of the tiling region because the second square of the domino is already in the bulk of the tiling region.

The start and end points of the line ensembles are moved accordingly. The LGV graph constructed in this way differs from the 6-vertex one by extra half-diagonal pieces added ($I_2,I_3$) or removed ($E_1,E_4$) which have unit weight. The other differences between the LGV graphs consists only in edges which, due to the boundary conditions, cannot be visited by the lines ($I_1,I_4,E_2,E_3$). Therefore the allowed line configurations are in one-to-one correspondence and have the same weight.

\subsubsection*{Particular case: DWBC and Aztec diamond}
A particular case is the six-vertex model in a square domain with domain wall boundary conditions, which means that a line starts from each SE point of $\partial S$ and a line ends at each SW point of $\partial S$. By the above mapping, the resulting domain to be tiled by dominos is the Aztec diamond, see Figure~\ref{FigDWBCaztecLines} for an illustration. This mapping between line ensembles can be easily inverted.

\subsubsection*{On more general boundary conditions}
Above we only established the mapping from the six-vertex model in a rectangular region $S$ to a domino tiling. However the construction holds for general domains provided the following four boundary configurations do not appear.
\begin{center}
\includegraphics[width=8cm]{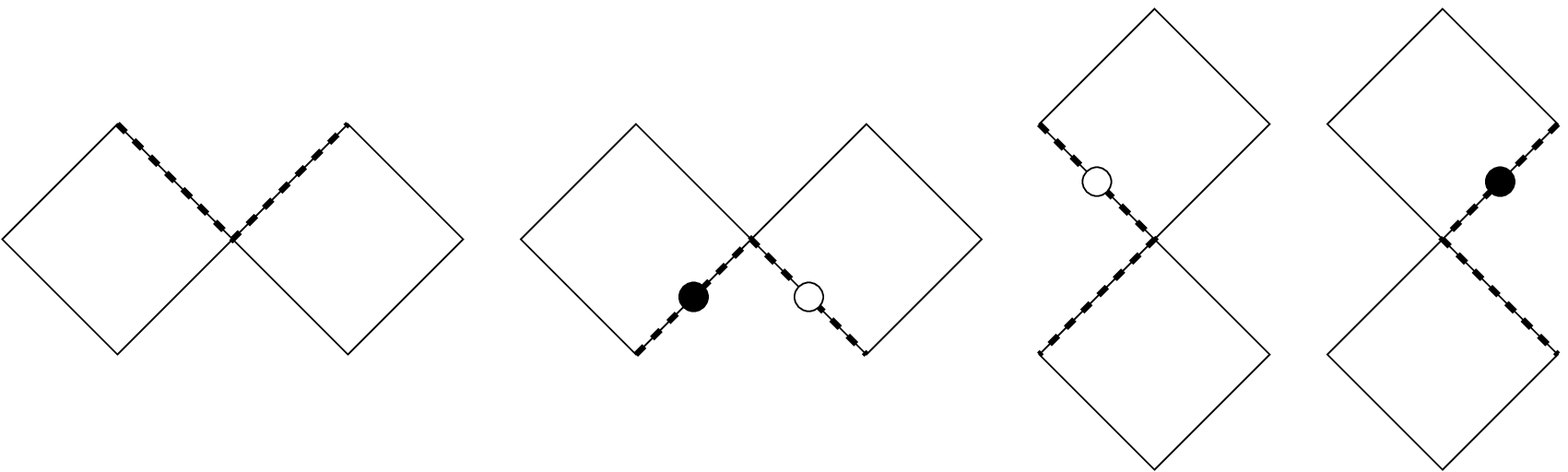}
\end{center}
The reason is that the above configurations are mapped into
\begin{center}
\includegraphics[width=8cm]{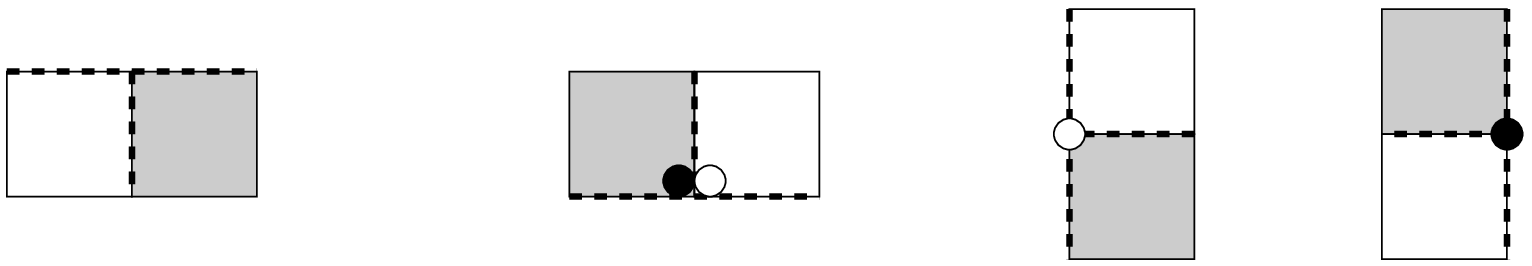}
\end{center}
and the natural boundary of the tiling no longer coincides with the dashed line.

\section{Conclusions and Outlook}
The top LGV line of the Aztec diamond borders, up to errors of size $1$, the North polar region, i.e., the region containing only North-type dominos which is connected with the North border of the Aztec diamond. For size $N$, let us denote the top LGV line by $h_N(t)$, $t\in[0,2N]$ (with the axis origin as in Figure~\ref{FigDWBCaztec}(b)). It follows immediately from the results by Johansson~\cite{Jo03} that $h_N(t)$ properly rescaled has limit as a random function. More precisely, let $\alpha=1$, then
\begin{equation}\label{eq1}
{\cal A}_N(\tau)=2^{5/6}N^{-1/3}\big(h_N(N+\tau N^{2/3}2^{-1/6})-N/\sqrt{2}\big)+\tau^2.
\end{equation}
Then ${\cal A}_N$ converges weakly to the process Airy process, $\tau\mapsto {\cal A}(\tau)$,
\begin{equation}\label{eq2}
\lim_{N\to\infty} {\cal A}_N(\tau)={\cal A}(\tau).
\end{equation}
The same limit process holds for all $\alpha>0$ and at all parts of the top LGV line dividing the North polar region and the disordered one. Of course, the scaling coefficients will change, but the scaling exponent and scaling function remain the same.

The result (\ref{eq1}), (\ref{eq2}) transcribes to the six-vertex model with domain wall boundary conditions. In the geometry of Figure~\ref{FigDWBCaztecLines} we may consider the outermost six-vertex line denoted by $h_{{\rm DW},N}(t)$. It borders the region tiled errorless with tiles of type $1$. Clearly, for all $t$ and $N$, one has
\begin{equation}
\big| h_N(t)-h_{{\rm DW},N}(t)\big|\leq 1.
\end{equation}
Therefore the limit (\ref{eq1}), (\ref{eq2}) holds also for the free-fermion six-vertex model.

\begin{figure}[t!]
\begin{center}
\psfrag{a}[c][]{$a$}
\psfrag{b}[c][]{$b$}
\psfrag{D}[c][]{$D$}
\psfrag{AF}[c][]{$AF$}
\psfrag{F}[c][]{$F$}
\includegraphics[height=4cm]{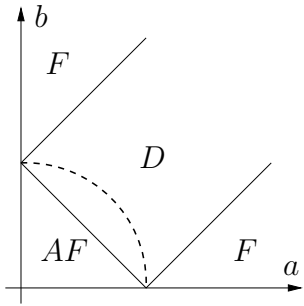}
\caption{The phase diagram for the six-vertex model. $F$ are the ferroelectric, $AF$ the anti-ferroelectric, and $D$ the disordered phases. The dashed line inside the disordered phase is the free fermion line, $\Delta=0$.}\label{FigPhaseDiag}
\end{center}
\end{figure}

The reader might be curious to know about the properties of the outermost line once the free fermion condition is no longer satisfied. For easier comparison with the literature, we switch to the standard parametrization
\begin{equation}
\omega_1=\omega_2=b,\quad \omega_3=\omega_4=a,\quad \omega_5=\omega_6=c
\end{equation}
with $a,b>0$ and exploiting the normalization of the weight we can set $c=1$.
If Figure~\ref{FigPhaseDiag} we display the bulk phase diagram for periodic boundary conditions and zero electric fields~\cite{LW72,ZJ02}. For DWBC one obtains the same phase diagram, however with a very different interpretation.
The phase diagram is governed by the parameter
\begin{equation}
\Delta=\frac{a^2+b^2-1}{2ab}.
\end{equation}
$\Delta=0$ is the free fermion line discussed before. It lies inside the disordered phase characterized by $-1<\Delta<1$.
$\Delta>1$ is the ferroelectric phase, while $\Delta<-1$ is the anti-ferroelectric phase. To understand the structure of typical configurations one has to resort to Monte Carlo simulations~\cite{SZ04,AR05}. For some special points in the phase diagram also analytical results are available.

In the disordered phase the structure is similar to the one at the free fermion line. There are then the perfectly ordered zones at the four corners with tiles $\omega_1,\omega_4,\omega_2,\omega_3$ counted clockwise starting from the top and a random zone which touches the border of the quadrant at four points. Presumably the macroscopic height profile satisfies the Pokrovsky-Talapov law, which is a strong indication that the outermost line is still governed by the Airy process (see~\cite{FPS03} for a discussion). As one crosses to the anti-ferroelectric phase, $\Delta<-1$, the structure of the outer configurations does not change qualitatively. However, the macroscopic height profile develops a facet at the center which consists of a regular arrangements of tiles $\omega_5$ and $\omega_6$. In contrast to the ordered zones at the corner, this facet allows for small statistical errors in the tiling. Thus it is less clear whether its (approximate) border ledge is still governed by the Airy process.

In the ferroelectric phase, $\Delta>1$, we choose $b>a$. Then the region below the anti-diagonal is filled with tiles of type $\omega_2$, while the one above with tiles of type $\omega_1$, separated by a rather sharp interface. The Monte Carlo simulations in~\cite{SZ04} indicate that the interface between these two perfectly ordered regions has a width which, as $N\to\infty$, goes to zero on a macroscopic scale, i.e., the width is $o(N)$.
If $b<a$, then the interface is along the diagonal, separating tiles of type $\omega_3$ from those of type $\omega_4$.

\subsection*{Acknowledgment}
We thank Michael Pr\"ahofer for illuminating discussions.

%\newcommand{\bibliodir}[1]{../../Biblio/#1}
%\bibliographystyle{\bibliodir{patplain}}
%\bibliography{\bibliodir{Biblio}}

\end{document}